# Gamma-Ray Transfer and Energy Deposition in Supernovae

Douglas A. Swartz[1][2]
Astrophysics Branch, ES-84
NASA Marshall Space Flight Center, Huntsville, AL 35812

Peter G. Sutherland
Department of Physics and Astronomy
McMaster University, Hamilton, Ontario, Canada L8S 4M1

Robert P. Harkness
University of Texas System Center for High Performance Computing
Balcones Research Center, Austin, TX 78758

## ABSTRACT

Solutions to the energy-independent (gray) radiative transfer equations are compared to results of Monte Carlo simulations of the $^{56}$Ni and $^{56}$Co radioactive decay $\gamma$-ray energy deposition in supernovae. The comparison shows that an effective, purely absorptive, gray opacity, $\kappa_\gamma \sim (0.06 \pm 0.01) Y_e$ cm$^2$ g$^{-1}$, where $Y_e$ is the total number of electrons per baryon, accurately describes the interaction of $\gamma$-rays with the cool supernova gas and the local $\gamma$-ray energy deposition within the gas. The nature of the $\gamma$-ray interaction process (dominated by Compton scattering in the relativistic regime) creates a weak dependence of $\kappa_\gamma$ on the optical thickness of the (spherically symmetric) supernova atmosphere: The maximum value of $\kappa_\gamma$ applies during optically thick conditions when individual $\gamma$-rays undergo multiple scattering encounters and the lower bound is reached at the phase characterized by a total Thomson optical depth to the center of the atmosphere $\tau_e \lesssim 1$. However, the constant asymptotic value, $\kappa_\gamma = 0.050 Y_e$ cm$^2$ g$^{-1}$, reproduces the thermal light curve due to $\gamma$-ray deposition for Type Ia supernova models to within 10% for the epoch from maximum light to $t = 1200$ days. Our results quantitatively confirm that the quick and efficient solution to the gray transfer problem provides an accurate representation of $\gamma$-ray energy deposition for a broad range of supernova conditions.

## 1. Introduction

Explosive silicon burning at high temperatures and densities during supernova explosions can produce unstable isotopes of iron group elements (Truran, Arnett &

---

[1]NAS/NRC Resident Research Associate

[2]Present Address: Nichols Research Corporation, 4040 S. Memorial Parkway, Huntsville AL 35815

Cameron 1967; Bodansky, Clayton & Fowler 1968). The decay of these isotopes and subsequent thermalization of the decay products can generate the energy needed to power the observed supernova optical display (Baade et al. 1956, Colgate & White 1966, Colgate & McKee 1969, Weaver & Woosley 1984) especially during the slowly evolving later phases. Numerical simulations invoking radioactive decay as a source of energy are able to reproduce observed supernovae light curves in detail. Not until SN 1987A, however, has the radioactive scenario been directly confirmed by the detection of $\gamma$-ray lines from $^{56}$Co decay (Matz et al. 1988; Arnett et al. 1989; Palmer et al. 1993) though indirect evidence for the fresh synthesis of $^{56}$Ni has been observed in optical (Axelrod 1980) and infrared (Varani et al. 1990) spectra by measurements of the evolution of thermally-excited Co emission features consistent with $^{56}$Co decay.

A practical consideration in supernova dynamics modeling is determining the local rate of radioactive energy deposition (or local heating rate). This rate provides the source term in the energy equation as part of the system of equations describing the physical state of the atmosphere. It can also be used directly in certain approximations for estimating rates of ionization and excitation by energetic electrons in statistical equilibrium (Axelrod 1980; Chugai 1987; Swartz 1991; Xu & McCray 1991; Kozma & Fransson 1992) and non-equilibrium (Fransson & Kozma 1993) equations. The radioactive source distribution can be provided by nucleosynthetic yields from explosion simulations, but the local heating rate is determined by the rate at which $\gamma$-rays and positrons deposit energy at various points in the gas as they travel through the atmosphere. (It is customary to assume that positrons will annihilate in the immediate vicinity of where they are produced. This is justified in terms of the relatively shorter range of electrons and positrons when compared with $\gamma$-rays and the likelihood of a tangled magnetic field sufficiently strong to inhibit the diffusion of charged particles, though see below.) Thus, in principle, a set of $\gamma$-ray transport equations must be solved simultaneously with the equations of hydrodynamics and equations of state in order to estimate observable quantities from models of supernovae. (The hydrodynamical evolution is often simplified by assuming homologous expansion which is justifiable once the ejecta have increased in radius by a factor of $\sim 10$ over the presupernova stellar size.)

Examples of the various methods for treating the $\gamma$-ray transport and energy deposition include Monte Carlo techniques (Colgate, Petschek & Kriese 1980; Ambwani & Sutherland 1988, hereafter AS; The, Burrows & Bussard 1990), multi-group (Weaver, Axelrod & Woosley 1980), and gray (Sutherland & Wheeler 1984, hereafter SW) radiative transfer methods as well as analytic approximations (Colgate, Petschek & Kriese 1980; Weaver, Axelrod & Woosley 1980; Arnett 1979; Ensman & Woosley 1988). It has been argued (SW) that the complex, multiple-scattering character of radiative transfer in supernova envelopes (where Compton scattering dominates) can, for the purposes of calculating energy deposition, be approximated by purely absorptive transfer. The plausibility of this assumption hinges on the following features of Compton scattering: (i) the scattering cross-section is forward-peaked, but forward-scattering leads to little energy transfer to the gas, and (ii) when a large-angle scattering takes place, there is significant energy transfer:

$$E'_\gamma = E_\gamma/(1 + (E_\gamma/m_e c^2)(1 - \cos\theta))$$

where $E_\gamma$ is the initial photon energy, $E'_\gamma$ is the energy after the photon scatters from an electron (mass $m_e$) at rest through angle $\theta$. Thus a $\sim 2$ MeV $\gamma$-ray (a representative energy for $^{56}$Ni and $^{56}$Co decay) gives up $\gtrsim 0.8$ of its energy if it scatters through $\theta \gtrsim 90°$. Crudely, then, one can think of the photon as proceeding along a nearly rectilinear path until it suffers a large-angle scattering, thereby giving up most of its energy. Ultimately, the credibility of the pure absorption approach depends on its producing results that are reasonably close to those from more realistic (i.e. Monte Carlo) simulations.

The purpose of the present work is to compare the results from the physically accurate but computationally intensive Monte Carlo simulations to solutions of transfer equations assuming a purely absorptive, energy-independent, opacity. Section 2 outlines the Monte Carlo and radiative transfer methods and the best fit value of the $\gamma$-ray opacity is determined in § 3. We find the purely absorptive effective opacity, $\kappa_\gamma$, can be expressed as $\kappa_\gamma = \kappa_\gamma^0 Y_e$ where $Y_e$ is the (total) number of electrons per baryon. In this form, $\kappa_\gamma$ is independent of the radioactive source distribution, mass density, and elemental abundance distributions (except for the dependence on abundances through $Y_e$). We find a weak dependence of $\kappa_\gamma$ on the total electron scattering optical depth, $\tau_e$, that results from a transition from a multiple to a single scattering environment encountered by the $\gamma$-rays in the Monte Carlo simulations. The best fit value of $\kappa_\gamma$ decreases from $\kappa_\gamma \sim (0.065 \pm 0.005)Y_e$ cm$^2$ g$^{-1}$ at $\tau_e \gtrsim 10$ to $\kappa_\gamma \sim 0.050 Y_e$ cm$^2$ g$^{-1}$ asymptotically for $\tau_e \lesssim 1$. A constant value for $\kappa_\gamma$ of $0.050 Y_e$ reproduces the Monte Carlo deposition and Type Ia light curves to within 10%.

By combining the effective opacity with a known radioactive source distribution in a supernova dynamic model, the relevant transfer equation becomes a linear first-order differential equation with constant coefficients and can be reduced to quadrature. Similarly, the Schwarzschild-Milne equations can be used to obtain the moments of the radiation field and hence the local deposition rate. In most practical applications, however, numerical integrations are required because the source function cannot be readily expressed analytically. Our radiative transfer algorithm to determine the local deposition function from known mass density and source distributions in spherical symmetry is available upon request to pgs@snowdrop.physics.mcmaster.ca. The processing time requirement for this algorithm (typically $\sim 0.05$ CPU seconds for a workstation for the full deposition function calculation) is a small fraction ($\sim 10^{-5}$) of the time required for a typical Monte Carlo simulation following 500,000 decay events.

## 2. Numerical Methods

The most abundant radioactive isotope produced in supernovae is $^{56}$Ni which decays with a 6.1 day half-life to $^{56}$Co which in turn decays with a 77.7 day half-life to $^{56}$Fe. The decays produce energetic $\gamma$-rays and positrons with initial energies 0.158 to 3.640 MeV (Lederer & Shirley 1978; AS). Supernovae yield other, less abundant, radioactive isotopes with longer life-times including $^{57}$Co, $^{44}$Ti, and $^{22}$Na. Typical photon energies for these sources range from $\lesssim 100$ keV to $\sim 1.3$ MeV. These long-lived isotopes become important only at very late times, $t > 1000$ days after the explosion (e.g., Woosley, Pinto, & Hartmann 1989), and are not considered in this work. One of the more unique aspects of the radioactive decay model for supernovae is that the ambient plasma remains cool ($T_e \sim 1$ eV) and nearly neutral while the primary heat source is characterized by photons that are six orders of magnitude more energetic. This can be seen by equating the energy density due to the slow radioactive energy release to that of an equivalent blackbody: Assuming the gas is optically thick to $\gamma$-rays then the mean intensity, $J \sim \epsilon_{rad}/4\pi\kappa_\gamma$, as shown below, where $\epsilon_{rad} \sim 7 \times 10^9 \exp(-t/\tau_{Co})$ ergs-s$^{-1}$-g$^{-1}$ is the energy generation rate per gram at times $t \gg \tau_{Ni}$ ($\tau_{Ni} = 7.58 \times 10^5$s and $\tau_{Co} = 9.80 \times 10^6$s are the $^{56}$Ni and $^{56}$Co decay times [AS]) and $\kappa_\gamma \sim 0.025$ cm$^2$-g$^{-1}$. This produces an energy density $\epsilon = 4\pi J/c \sim \epsilon_{rad}/c\kappa_\gamma$ equivalent to that of a blackbody at $T \sim (\pi J/\sigma)^{1/4} \sim 6000 \exp(-t/4\tau_{Co})$ K where $\sigma$ is the Stefan-Boltzmann constant.

The decay photons primarily interact with the surrounding gas through Compton scattering. A fraction of the photon energy is lost per scattering and the down scattered photon is typically destroyed in K-shell photoionization in less than five scatterings (AS). These interactions are independent of the temperature and ionization state of the gas at the low-temperature near-neutral conditions typical of supernovae. The decay positrons and recoil electrons are the primary ionizing and heating agents in the process. Their range is only a fraction of the distance across the atmosphere and their energy loss is typically treated *in situ* while the $\gamma$-rays can travel long distances and often escape the atmosphere altogether. Thus the problem of calculating the local energy deposition rate is equivalent to calculating the rate of production of electrons and positrons and their initial spectrum. Positron decays account for 19% of all $^{56}$Co decays. The average decay positron energy is $\sim 0.66$ MeV (0.125 MeV per decay or $\sim 4\%$ of the total decay energy) and the maximum decay energy is 2.46 MeV. This kinetic energy is not accounted for in the Monte Carlo energy deposition calculations although the two annihilation photons, each of energy $E_\gamma = m_e c^2$, are followed. If the kinetic energy is deposited very near the site of the decay, then a small additional energy deposition term, proportional to the local mass fraction of radioactive $^{56}$Ni, should be added to the results from either the Monte Carlo calculation or the gray absorption calculations which follow. If the assumption that charged particles are trapped locally is invalid, as has been suggested again recently by Chan & Lingenfelter (1993), then *neither* the Monte Carlo nor gray absorption calculations considered here are appropriate. Chan & Lingenfelter (1993) argue that radially-combed magnetic fields may facilitate the escape of the positrons and other charged particles. But then at most epochs a very large fraction of the radioactive decay energy will escape and this would appear to be in conflict with our understanding of any supernovae whose light curves we believe to be powered by radioactivity.

## 2.1. Monte Carlo Simulations

The Monte Carlo code used in this work is essentially that described by AS and by Sutherland (1990). This code was originally developed to calculate the emergent $\gamma$-ray and X-ray spectra and incorporated Doppler effects associated with the propagation of the photons through a homologously expanding supernova atmosphere. The present version includes the complete $^{56}$Ni and $^{56}$Co decay source spectrum (AS) and interactions with the gas through Compton scattering, pair production, and photoelectric absorption. Compton scattering is described by the Klein-Nishina cross-section and all electrons, bound and ionized, are included. Cross sections for the other processes are given in AS with the following correction: The pair production cross section, $\sigma_{pair}$, per electron is

$$\sigma_{pair} = \frac{\sigma_{pair}^0}{n_e} \sum_i n_i Z_i^2 \qquad (1)$$

where

$$\sigma_{pair}^0 = \begin{cases} 0 & E_\gamma < 2m_e c^2 \\ 0.10063(E_\gamma - 1.022) & E_\gamma < 1.5 \text{ MeV} \\ 0.0481 + 0.301(E_\gamma - 1.5) & E_\gamma > 1.5 \text{ MeV} \end{cases} \qquad (2)$$

in units of $10^{-27}$ cm$^2$/atom. Here, $E_\gamma$ is the photon energy in MeV, $n_e$ is the total electron density, and $n_i$ is the number density of element $i$ with nuclear charge $Z_i$. (Compare equation 2 of AS.)

The prescription for photoelectric absorption in this code is the following:

$$\sigma_{pe} = (E_\gamma/100 \text{ keV})^{-3} \sum_i \frac{n_i}{n_e} \sigma_{pe,i}^{100} \equiv \sigma_{pe}^0 \left(E_\gamma/100 \text{ keV}\right)^{-3} \qquad (3)$$

per electron where $\sigma_{pe,i}^{100}$ is the photoelectric cross-section (per atom) for the $i^{th}$ element evaluated at 100 keV and the sum is over all ion species. The cross-section data are from the tables of Veigele (1973). Numerical values for $\sigma_{pe}^0$ for the several compositions used in this work are given in Table 1. These cross-sections are for neutral atoms but are dominated by the effects of the innermost electrons which are not expected to be ionized, except for H, in the conditions of interest for SNe ejecta. The scaling $E^{-3}$ for photoelectric absorption is an excellent approximation (especially between 10 and 1000 keV), but in any case the details are less important for deposition calculations because by the time a $\gamma$-ray is likely to suffer photoelectric absorption its energy has been so degraded (to $\leq 30$ keV) by repeated Compton scattering that the impact upon energy deposition is small. (On the other hand, in calculating emergent X-ray spectra, it is important to describe accurately the processes of photoelectric absorption and fluorescence.)

The energy deposited locally by any of the above processes is then: (i) the energy difference between the incident and scattered photon, or (ii) the energy difference between the incident photon and $2m_e c^2$ if a pair is produced (with the creation of two 511 keV $\gamma$-rays), or (iii) the full energy of the photon if it is photoelectrically absorbed. The local deposition function is defined as the ratio of the local energy deposition rate per unit mass to the rate of energy generation per unit mass of radioactive material. For a simulation following the fate of $N$ decay events, the value of the deposition function in the $k^{th}$ (Lagrangian) mass shell is the product of (i) the ratio of the total energy deposited in the shell to the total energy of decay photons ($\sum_\gamma N f_\gamma E_\gamma$, where photons of energy $E_\gamma$ are distributed according to the probability of emission, $f_\gamma$, as tabulated in, e.g., AS) and (ii) the weight $M_{rad}/\Delta M_k$ where $M_{rad}$ is the total mass of initially radioactive material and $\Delta M_k$ is the mass of the shell. This definition has the property that at early epochs, when the ejecta are optically thick and essentially all the $\gamma$-rays are trapped near where they are emitted, the local deposition function equals the local mass fraction of radioactive material.

Typical Monte Carlo simulations involve a minimum of 500,000 decay events and models with 64 zones. Comparison with simulations with even more decays confirmed that this number of decays ensures errors less than 5% for all zones for all models.

A separate code that assumes a purely absorptive interaction of photons with matter was also implemented, and in this case the atmosphere was treated as static so that comparison could be directly made with the gray transfer approach of the next section.

## 2.2. Gray Radiative Transfer

The transfer equation in spherical symmetry is cast in a difference form and integrated along impact parameters parallel to an observer's line of sight. We assume the $\gamma$-ray opacity, $\kappa_\gamma$, is independent of energy, $E$, and purely absorptive so that the transfer equation for the energy-integrated intensity, $I \equiv \int_o^\infty I_E dE$, along a ray is

$$\pm \frac{\partial I^\pm}{\partial z} = \eta - \kappa_\gamma \rho I^\pm \tag{4}$$

for the incoming ($I^-$) and outgoing ($I^+$) directions where $z$ denotes the position along the ray, $\rho$ is the mass density, and $\eta \equiv \int_o^\infty \eta_E dE$ is the local total $\gamma$-ray emissivity: $\eta = f_{rad}\epsilon_{rad}\rho/4\pi$. Here, $f_{rad}$ is the initial mass fraction of $^{56}$Ni and $\epsilon_{rad}$ denotes the time-dependent rate of energy release per gram of radioactive material. In cgs units,

$$\epsilon_{rad} = 3.9 \times 10^{10} \exp(-t/\tau_{Ni}) + 6.78 \times 10^9 \left[\exp(-t/\tau_{Co}) - \exp(-t/\tau_{Ni})\right]. \tag{5}$$

By introducing the optical depth along the ray,

$$d\tau = -\kappa_\gamma \rho dz, \tag{6}$$

and defining

$$I' = (4\pi\kappa_\gamma/\epsilon_{rad})I, \tag{7}$$

equation 4 can be written

$$\frac{dI'}{d\tau} = I' - f_{rad}. \tag{8}$$

Since no radiation is incident from outside the ejecta, $I^- = 0$ at the upper boundary and, from symmetry at $z = 0$, the lower boundary condition is $I^+ = I^-$. Integration of equation 8 yields

$$I'(\tau_i) = I'(\tau_{i+1})e^{-\Delta\tau} + \int_{\tau_i}^{\tau_{i+1}} f_{rad} e^{-(t-\tau_i)} dt \tag{9}$$

where $\tau_i$ and $\tau_{i+1}$ represent optical depth points along the ray and $\Delta\tau = (\tau_{i+1} - \tau_i)$. Since $f_{rad}$ is constant within each shell, equation 9 can be integrated *exactly*. The impact parameter grid is constructed so that all rays are tangent to radial shells with a single 'core ray' passing through the center.

The radiative transfer code has been tested successfully against another transfer code which casts the transfer equation in a second-order difference form and uses a Feautrier solution along impact parameters. From this solution, variable Eddington factors can be evaluated and used in a combined moment equation constructed using Auer's transformation, again employing a Feautrier solution. This latter code has been extensively tested to reproduce various analytic results including the proper asymptotic results in the gray case (e.g. Hummer & Rybicki 1971).

The rate at which energy is deposited locally (in ergs cm$^{-3}$ s$^{-1}$) is $4\pi\kappa_\gamma\rho J$ where $J$ is the mean intensity:

$$J \equiv \frac{1}{4\pi} \oint I d\omega \qquad (10)$$

which is the specific intensity integrated over all solid angles $d\omega$. Thus the deposition function is:

$$d = \frac{4\pi\kappa_\gamma\rho J}{\epsilon_{rad}\rho} = \frac{4\pi\kappa_\gamma J}{\epsilon_{rad}}. \qquad (11)$$

The deposition function is (compare equations 11 and 7) then simply $J'$, the mean of $I'$.

The deposition function can be calculated using the spherically symmetric gray radiative transfer equations once the mass density $\rho$, initial $^{56}$Ni mass fraction $f_{rad}$, and the $\gamma$-ray opacity $\kappa_\gamma$ are specified on a radial grid. The calculation does not require knowledge of the source spectrum but only of the total rate of emission, $\epsilon_{rad}$.

## 3. Results

The Monte Carlo solution is the standard by which the deposition function calculated using the gray radiative transfer method is to be compared. The deposition function is determined by the density profile and the distribution of the radioactive material and is therefore position dependent. A global measure describing the goodness of fit of the gray results to the Monte Carlo results must take into account this spatial dependence. Assuming the atmosphere is composed of $ND$ concentric shells of radii $r_k$, $k = 1, ND$, the quality of fit, $Q$, is defined to be

$$Q = \sum_{k=1}^{ND} w_k \left(\frac{d_k - d_k^g}{\sigma_k}\right)^2 \qquad (12)$$

where $d$ and $d^g$ are the local values of the deposition function calculated using the Monte Carlo and the gray radiative transfer methods, respectively. The uncertainties $\sigma_k$ are those calculated per zone in the Monte Carlo calculation and in effect reflect "counting statistics" (they are obtained by accumulating the variance for each zone in the energy deposited by those interactions that occurred in each zone — $\sigma_k$ is related to the "error in the mean" for the deposited energy). We include a weight factor $w_k$ for each zone to compensate for the following (if $w_k = 1$ then $Q$ is essentially a $\chi^2_{ND}$ statistic): First, we wish to downplay low mass zones that were used in the zoning to achieve resolution for either the density or velocity profiles. Secondly, at times when it is of most interest to calculate the deposition function, the thermal diffusion timescale in the ejecta is short compared with the dynamic timescale. Then the luminosity is given by the instantaneous balance between $\gamma$-ray heating and radiative cooling: $L = \sum_k \Delta M_k d_k \epsilon_{rad}$ so that regions of small $d$ do not contribute significantly to the observed luminosity. Thus $Q$ should be weighted to emphasize those zones which contribute most to the luminosity. Thus we have chosen the weights $w_k = \Delta M_k / M_{tot}$. The gray deposition, $d_k^g$, is a function of the single parameter, $\kappa_\gamma$. Therefore, the optimum value of $\kappa_\gamma$ is determined for each Monte Carlo simulation by minimizing $Q$ in equation 12.

For purposes of exposition reference will be made to the net deposition (from which the luminosity, $L$, follows trivially) defined as

$$D_\gamma = \sum_{k=1}^{ND} \frac{d_k \Delta M_k}{M_{rad}} \qquad (13)$$

where $M_{rad} \equiv \sum_k \Delta M_k f_{rad,k}$. The net deposition has the property that $D_\gamma \to 1$ as $\tau \to \infty$ and $D_\gamma \to 0$ as $\tau \to 0$.

Simulations were performed for a range of plausible supernova conditions including various mass and source distributions, (homogeneous) elemental abundances, and $\gamma$-ray optical depths. Two basic models and their time evolution were considered. The first model is based on the incinerated white dwarf model W7 of Nomoto, Thielemann & Yokoi (1984) that is known to provide a good fit to the observed light curves (Nomoto, Thielemann & Yokoi 1984) and spectra (Harkness 1991) of typical Type Ia supernovae. The total ejected mass for this model is $\sim 1.4$ M$_\odot$, there is $\sim 0.6$ M$_\odot$ of radioactive material within the inner $\sim 1$ M$_\odot$ with a peak in this $^{56}$Ni distribution near $M_r \sim 0.7$ (where $M_r$ is the mass interior to the point $r$). The central regions of the model, $M_r \lesssim 0.1$ M$_\odot$, are depleted of radioactive material during neutron-rich freezeout, and the outer $\sim 0.4$ M$_\odot$ is composed of partially incinerated, intermediate mass elements that result from carbon-burning. The detailed abundances are replaced with mass-weighted mean values in the present work, but the original $^{56}$Ni distribution is retained intact.

The second model assumes a power law density profile ($\rho \propto r^{-\alpha}$) with uniform abundances. We examined models with total masses in the range 1 to 5 M$_\odot$ and power law exponents, $\alpha$, in the range 0 (uniform sphere) to 5 with outer radius $R = v_{max} t$, $v_{max} \sim 1.0 \times 10^9$ cm s$^{-1}$. Both a uniform $^{56}$Ni distribution and a central source distribution were considered in the power law models. For the central source, the radioactive source distribution is represented by a step function with the inner 1% of the total mass assumed radioactive. Chemical compositions for model W7 and for two representative mixtures are listed in Table 1. The solar composition represents conditions found in the outer layers of typical Type II supernovae and the mixture 10H represents Type II supernovae with all ejected material (core and envelope) homogenized. Each model is initialized using $ND = 64$ Lagrangian mass shells and scaled homologously in time.

### 3.1. The Effective $\gamma$-ray Opacity, $\kappa_\gamma$

It is instructive to determine the effective opacity for mono-energetic sources before considering the full $^{56}$Ni and $^{56}$Co decay spectrum. The cross sections for the interaction processes included in the Monte Carlo simulations are dominated by Compton scattering for photon energies spanning the $^{56}$Ni and $^{56}$Co decay spectral range, $0.158 < E_\gamma < 3.640$ MeV (Figure 1). The (angle-averaged) Klein-Nishina cross section changes monotonically by a factor of 4 within this energy band. The effective opacity encountered by a decay photon should therefore depend upon its initial energy. Figure 1 shows the best fit to the effective opacity (obtained by minimizing $Q$) for mono-energetic $\gamma$-rays as a function of the initial photon energy for model W7 and for several power law models at an intermediate stage of evolution, $\bar{D}_\gamma \sim 0.3$ and $\tau_e \gtrsim 3$ (see below). The composition for all models is the mass-averaged W7 model abundances. The effective opacity is seen to be independent of the model density profile and source distribution but varies approximately in proportion to the total cross section at $E_\gamma$ for initial photon energies between $\sim 0.2$ and $\sim 6$ MeV. The effective opacity is between 0.5 and 0.7 of the total opacity in this energy range. The effect of the purely absorptive photoionization process causes this fraction to rise at lower energies with the effective opacity approaching the photoionization opacity at energies $E_\gamma \lesssim 0.1$ MeV, depending on the relative contribution of photoionization to the total opacity.

The Monte Carlo deposition function is shown for several initial $\gamma$-ray energies against the Lagrangian mass, $M_r$, in Figure 2 for a central source in a 1 $M_\odot$ uniform density model. This illustrates the increasing opacity for lower initial energy $\gamma$-rays from another perspective. The deposition functions for initial energies $E_\gamma > 2$ MeV are nearly identical reflecting the similar effective opacities in this energy range. At lower initial photon energies, the deposition tends to accumulate near the point of emission reflecting a relatively higher opacity. Since each interaction either destroys the photon or emits a scattered photon at a reduced energy, the opacity increases following every scattering event. Thus the effective opacity, which reflects the changing opacity seen by the photon as it scatters and loses energy, is larger than the absorptive opacity experienced only by photons at a single initial energy. Further, a photon that has lost a significant amount of energy in a scattering event will have a high probability of another scattering or absorption event nearby.

The fact that the effective opacity depends on the initial photon energy suggests that the effective opacity will be time dependent for two reasons. First, the photon energy distribution changes from a $^{56}$Ni-dominated spectrum to a $^{56}$Co-dominated spectrum as the supernova evolves (Figure 1 and Table 1 of AS). The effective opacity for a pure $^{56}$Ni source can be expected to differ from that for a pure $^{56}$Co source (and from that for a $^{57}$Co or other radioactive source). These differences are, however, negligible compared to the effect due to expansion of the atmosphere which reduces the optical thickness and hence the probability that a photon will interact with the gas.

The time dependence (represented by the total electron scattering optical depth) of the best fit effective opacity for a mono-energetic source is shown in Figure 3 for three of the models from Figure 1 ($Y_e = 0.5$). The effective opacity evolves with changing optical depth but is seen to be independent of the model source and mass distributions. (The error bars indicate 1-$\sigma$ confidence intervals for $\kappa_\gamma$. Since these errors are determined by the number of decay events followed in the Monte Carlo simulations, they are shown only to indicate the *relative* errors and are chosen conservatively for clarity.) A general trend from large $\kappa_\gamma$ at high $\tau_e$ to small $\kappa_\gamma$ at low $\tau_e$ is evident. Above $\tau_e \sim 1$, $\kappa_\gamma$ has a weak power law dependence on $\tau_e$. At the highest optical depths the destruction length is short and nearly all the $\gamma$-rays are thermalized near their points of origin. In this limit, $\tau \to \infty$ and $D_\gamma \to 1$, the deposition functions are nearly identical for a range of $\kappa_\gamma$ values above $\sim 0.03$. This is also evident from equations 9 – 11 where, in the optically thin limit, $d \to f_{rad}$ independent of the value of $\kappa_\gamma$. In the opposite limit, $\tau \ll 1$, $\kappa_\gamma \to 0.028$ cm$^2$ g$^{-1}$ for photons with initial energy $E_\gamma$ = 1 MeV. This dependence on $\tau_e$ can be understood qualitatively as follows: Suppose at some epoch for which $\tau_1 \sim 1$ ($\tau_1$ is the center to surface optical depth for the scattering of, for example, a 1 MeV photon through 90 degrees or more) the best fit between the Monte Carlo and gray calculations is achieved for $\kappa_1 = \kappa_1^0$. At this epoch, some gamma-rays escape without depositing energy, and in the Monte Carlo calculation even those photons that scatter will not give up all their energy. The combination of these two effects will determine $\kappa_1^0$. However, at earlier epochs, when $\tau_1 >> 1$ very few photons in the Monte Carlo calculation will escape, and by virtue of repeated scatterings will give up virtually all their energy. Thus $\kappa_1$ will have to be increased to compensate. Conversely, at epochs for which $\tau_1 << 1$, the "real" Monte Carlo photons rarely scatter more than once, almost never give up their entire energy, and $\kappa_1$ must accordingly be reduced below $\kappa_1^0$. Put more succinctly, the effect of multiple scatterings is to increase the "efficiency" of energy deposition and therefore at earlier epochs $\kappa_1$ must be slightly increased. That $\kappa_1$ approaches 0.028 cm$^2$ g$^{-1}$ as $\tau_e \to 0$ can be confirmed for a central point source with the following simple numerical calculation: The effective cross-section for single-scattering energy deposition by a $\gamma$-ray line of energy $0.2 \lesssim E_\gamma \lesssim 10$ MeV is dominated by the energy-loss-weighted Klein-Nishina cross-section:

$$\sigma_{eff}(E_\gamma) = (1/E_\gamma) \int \frac{d\sigma_{KN}(E_\gamma)}{d\Omega} \Delta E d\Omega \quad (14)$$

where the energy loss is given by $\Delta E/E_\gamma = (E_\gamma/m_e c^2)(1-\cos\theta)/(1+(E_\gamma/m_e c^2)(1-cos\theta))$. For $E_\gamma = 1$ MeV, $\sigma_{eff} = 0.14$ in units of the Thomson cross-section ($\kappa_{eff} = 0.028$ cm$^2$ g$^{-1}$ for $Y_e = 0.5$). This result has been confirmed for an optically thin sphere with central point source, where only single-scattering is important and the optical depth to the surface is isotropic. This argument must also extend to all optically-thin situations. Figure 4 displays $\sigma_{eff}$ for a range of $\gamma$-ray energies based on equation 14. Figure 4 also shows that the angle-averaged fractional energy loss is $\sim 0.5$ for $E_\gamma \gtrsim 1.0$ MeV indicating that about half the photon energy is lost per scattering, on average, for photons in this energy range. Thus, if only one scattering is encountered by a photon, we can expect an effective opacity $\kappa_\gamma \sim 0.5\kappa_{KN} = 0.5(\sigma_{KN} Y_e/m)$, in agreement with Figure 1. This does not account for photoionization which adds considerably to the effective opacity at low energies. The true effective opacity, $\kappa_\gamma$, which accounts for all interaction processes and all subsequent scatterings, exceeds the 'single-scattering' effective Compton opacity, $\kappa_{eff}$. Thus, $\kappa_{eff}$ is an approximate lower limit for $\kappa_\gamma$ in that $\kappa_\gamma \to \kappa_{eff}$ in the optically thin limit and for photon energies, $E_\gamma$, far above the photoionization threshold. This reasoning also implies, as confirmed by the Monte Carlo simulations, that the largest energy loss occurs during the first scattering event.

These arguments extend to the full $^{56}$Ni and $^{56}$Co decay spectrum. The time evolution of the effective opacity for several models is shown in Figure 5 using the full time-dependent decay spectrum. The trends noted above are reproduced by the full spectrum: the best fit values of $\kappa_\gamma$ are independent of the source and density distributions and $\kappa_\gamma$ has a weak dependence on $\tau_e$, decreasing from $\sim 0.035 \pm 0.01$ at $\tau_e \gtrsim 3$ to $\sim 0.025$ for $\tau_e < 1$. The effective pure absorption cross-section for single scattering (equation 14) is now:

$$\sigma_{abs} = \frac{\sum_\gamma f_\gamma E_\gamma \sigma_{eff}(E_\gamma)}{\sum_\gamma f_\gamma E_\gamma} \qquad (15)$$

where the sum is taken over the line spectrum ($f_\gamma$ is the probability for a given line). The result of doing the above calculation for the line spectrum of $^{56}$Co decay is $\sigma_{abs} \sim 0.128$ in units of the Thomson cross-section ($\kappa_\gamma = 0.051 Y_e$ cm$^2$ g$^{-1}$).

Figure 6 shows that $\kappa_\gamma$ for different compositions can be written as $\kappa_\gamma \equiv \kappa_\gamma^0 Y_e$ where $Y_e$ is the number of electrons per baryon (in all the preceding figures, $Y_e = 0.5$). This is the well-known Compton scattering dependence on the number of electrons: For Compton scattering, the effective opacity can be written as $\kappa_\gamma = (n_e \sigma_{KN}/\rho)$ where the total (bound and free) electron density is $n_e = (\rho/m) Y_e$ and $m$ denotes the atomic mass unit. Thus $\kappa_\gamma$ can be expressed as $\kappa_\gamma = (\sigma/m) Y_e$ or, equivalently, as $\kappa_\gamma \equiv \kappa_\gamma^0 Y_e$. The values of $\kappa_\gamma$ shown in Figure 6 and in previous figures are consistent with values quoted in the literature. For example, Colgate, Petschek, & Kriese (1980) found $\kappa_\gamma = 0.028$ cm$^2$ g$^{-1}$ for a pure Ni 0.5 M$_\odot$ uniform sphere Type Ia supernovae model. Woosley, Pinto, & Hartmann (1989) quote a value of $0.05 Y_e$ for the 10H composition model for SN 1987A at late times.

### 3.2. Deposition Functions and Light Curves

The quantity $Q$ provides a global statistical measure of the agreement of the gray deposition function to the results of the Monte Carlo simulation. By minimizing $Q$, the best-fit value of the parameter $\kappa_\gamma$ is determined. What is also needed is a measure of the accuracy of the results parameterized by $\kappa_\gamma$. Two physical properties that characterize the $\gamma$-ray transport phenomenon and that are fundamentally relevant to supernova studies are the local heating rate and the light curve. These are related to the local deposition function, $d_k$, and the net deposition, $D_\gamma$, respectively, and their time evolution. The accuracy of the gray radiative transfer results can be estimated by comparing the local values of the deposition functions and net deposition to the Monte Carlo results.

Figure 7 shows the local values of the deposition functions for model W7 at several times, $t$. The optimal value of $\kappa_\gamma$ at $t \sim 60$ days ($\tau_e \sim 12$) for this model is $\kappa_\gamma = 0.059 Y_e$. This value of $\kappa_\gamma$ reproduces the deposition function in all regions of the model at 60 days to within a few percent. At earlier times ($t \sim 35$ days) using this value of $\kappa_\gamma$ there is a 6% discrepancy in the net deposition, $D_\gamma$, computed using the two methods with the Monte Carlo deposition function higher than the gray deposition function by $\sim 6\%$ at the peak of the distribution. At even earlier times, the fit is actually better due to the degeneracy of $\kappa_\gamma$ values that produce a good fit (see above). At later times, $t = 90$ days, the deposition computed using the value $\kappa_\gamma = 0.059 Y_e$ results in a net deposition $\sim 6\%$ higher and local deposition functions $< 10\%$ larger than those computed using the Monte Carlo calculations. These results are indicative of those found for other models. Figure 8 shows that changing $\kappa_\gamma$ by $\sim 20\%$ from the best fit value at 60 days produces deposition functions that differ by at most 12% from the Monte Carlo results.

Supernovae are initially hot and sufficiently ionized to be optically thick. As the supernova debris expands, the atmosphere cools and the photosphere recedes to deeper layers. Eventually, as the thermal diffusion timescale becomes short compared to the dynamic timescale, the energy deposited by $\gamma$-rays is instantaneously balanced by thermal losses. During these late times an excellent approximation to the bolometric luminosity (not including, by convention, escaping decay $\gamma$-rays and down-scattered x-rays) can be made by simply equating the instantaneous $\gamma$-ray energy deposition rate to the bolometric luminosity. The light curve for model W7, defined as $L(t) = \sum_k \Delta M_k d_k(t) \epsilon_{rad} = \epsilon_{rad} D_\gamma(t) M_{rad}$, is shown in Figure 9 computed from the Monte Carlo simulation and from the (time-dependent) best fit opacity gray calculation. Light curves computed using time-independent values of $\kappa_\gamma = 0.02$, 0.025, and 0.03 ($Y_e = 0.5$) are also illustrated. For $D_\gamma \gtrsim 0.95$, the luminosities computed using either of these four values for $\kappa_\gamma$ are accurate to better than $\sim 3\%$. The light curve computed using the time-dependent best-fit $\kappa_\gamma$ is accurate to within 2% at all times. Those for $\kappa_\gamma = 0.02$, 0.025, and 0.03 are in error by less than 22%, 10%, and 18%, respectively. The largest error occurs at $t \sim 90$ days for $\kappa_\gamma = 0.02$ and 0.025 and at $t = 1200$ days for $\kappa_\gamma = 0.03$. It should be noted that in the computation of realistic light curves one would also include the kinetic energy of the positrons associated with 19% of the $^{56}$Co decays. If there are no radially-combed magnetic field lines that facilitate the escape of these positrons then the positron kinetic energy contribution would correspond to an additional, *local*, energy deposition term $\sim 0.04 \epsilon_{rad} * f_{rad}$ which should be added to the results from either the Monte Carlo calculation or the gray absorption calculation. As noted earlier, if the magnetic field is radially-combed, then the positrons and most of the primary, high-energy electrons (produced by first and second scatterings of $\gamma$-rays) will escape and the calculation of the energy deposition function will differ significantly from that described here.

## 4. Conclusions

The radioactive model for supernovae, in which freshly synthesized $^{56}$Ni and its daughter nucleus, $^{56}$Co, power the late-time luminosity, is well established and has been confirmed observationally. Numerical applications of the radioactive model are burdened by the complexity of the decay $\gamma$-ray interactions with the supernova material. Though physically accurate, traditional Monte Carlo techniques are excessively demanding of computer resources especially for light curve studies where the calculation of the $\gamma$-ray energy deposition must be repeated many times. Many alternative methods have been invoked yet none have been rigourously justified through quantitative comparisons to Monte Carlo methods.

Perhaps the simplest means of computing the $\gamma$-ray energy deposition, which is also computationally efficient and physically reasonable, is the gray radiative transfer method. The gray transfer technique was compared in detail to Monte Carlo simulations in this work. Solutions to the gray transfer equations can be matched to the corresponding Monte Carlo results by adjusting a single parameter: the energy-independent material opacity, $\kappa_\gamma$.

We applied a merit function, $Q$, similar to a $\chi^2$-statistic, to systematically identify the best-fit value for $\kappa_\gamma$ which was found to be weakly dependent on the optical thickness of the supernova atmosphere. This dependence can be explained qualitatively by the fact that as photons downscatter through successive interactions the opacity increases with the result that the effective opacity is larger for greater optical depths. This dependence also reflects the fact that the gray radiative transfer model is not fully appropriate to the $\gamma$-ray deposition problem. The accuracy of the gray transfer solution fits to the Monte Carlo deposition functions was also analyzed. Using the optimal values of $\kappa_\gamma$, obtained by minimizing $Q$, produces fits to a typical Type Ia supernova light curve due to $\gamma$-ray deposition that are accurate to within 2% of the Monte Carlo results for times $t$ from maximum light ($t \sim 12$ days) to the latest phase included in the study ($t = 1200$ days). Neglecting the weak optical depth dependence of $\kappa_\gamma$ by using the asymptotic value of $\kappa_\gamma$ = $0.05Y_e$ (the optimal value for $\tau_e \lesssim 1$) also produces an acceptable fit to the light curve with errors not exceeding 10%.

---

TABLE 1
Model Composition by Mass Fraction[1]

| Element | W7[2] | 10H[3] | Solar[4] |
|---|---|---|---|
| $^1$H |  | 3.1 (-1) | 7.7 (-1) |
| $^2$He | 8.2 (-3) | 5.1 (-1) | 2.1 (-1) |
| $^6$C | 9.3 (-5) | 1.5 (-2) | 3.8 (-3) |
| $^7$N |  | 2.5 (-4) | 9.3 (-4) |
| $^8$O | 1.0 (-1) | 9.1 (-2) | 8.5 (-3) |
| $^{10}$Ne | 3.1 (-4) | 3.9 (-3) | 1.5 (-3) |
| $^{12}$Mg | 1.9 (-2) | 1.1 (-3) | 7.4 (-4) |
| $^{13}$Al | 6.0 (-4) |  | 6.6 (-5) |
| $^{14}$Si | 1.3 (-1) | 1.8 (-2) | 8.1 (-4) |
| $^{15}$P | 1.5 (-4) |  | 5.8 (-6) |
| $^{16}$S | 6.6 (-2) | 1.1 (-2) | 4.6 (-4) |
| $^{17}$Cl | 1.3 (-4) |  | 4.8 (-6) |
| $^{18}$Ar | 1.4 (-2) | 2.0 (-3) | 1.2 (-4) |
| $^{19}$K | 8.1 (-5) |  | 3.9 (-6) |
| $^{20}$Ca | 1.1 (-2) | 1.2 (-3) | 7.2 (-5) |
| $^{22}$Ti | 2.7 (-5) | 2.5 (-5) | 3.3 (-6) |
| $^{24}$Cr | 7.0 (-4) | 3.4 (-5) | 1.9 (-5) |
| $^{25}$Mn | 1.3 (-4) |  | 1.5 (-5) |
| $^{26}$Fe | 8.0 (-2) | 5.2 (-4) | 1.5 (-3) |
| $^{27}$Co | 4.1 (-3) |  | 3.7 (-6) |
| $^{28}$Ni | 5.7 (-1) | 3.1 (-2) | 8.1 (-5) |
| $Y_e$ | 0.50 | 0.66 | 0.89 |

[1] power of ten exponent in parentheses
[2] from Thielemann et al. (1986) model W7
[3] from Woosley (1988) model 10H
[4] from Cameron (1982)
– 15 –

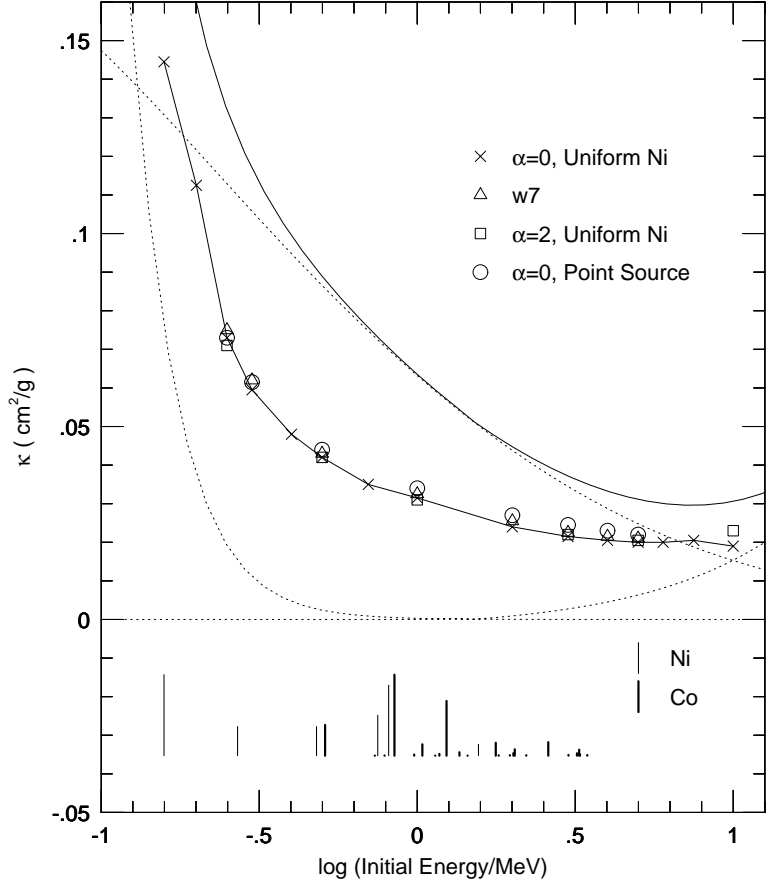

Fig. 1.— The best-fit effective opacity to mono-energetic sources for several models is shown against the initial photon energy. The uppermost solid line represents the total opacity which is composed of photoabsorption (*dotted line*), Compton scattering (*short-dashed line*), and pair production (*long-dashed line*). All models assume the W7 abundances ($Y_e = 0.5$) of Table 1, $\alpha$ denotes the power law exponent for power law models and model w7 is the white dwarf model of Nomoto, Thielemann, & Yokoi (1984). The total $\tau_{es}$ to the center of the ejecta for these models ranges from $\sim 3$ to 12. The lower portion of the figure schematically illustrates the $^{56}$Ni and $^{56}$Co decay spectra. The height of the vertical lines depicts the relative probabilities of the individual decay paths hence the strongest lines are the $^{56}$Ni 158 keV ($f_\gamma = 1.0$) and $^{56}$Co 847 keV ($f_\gamma = 0.9998$) lines.

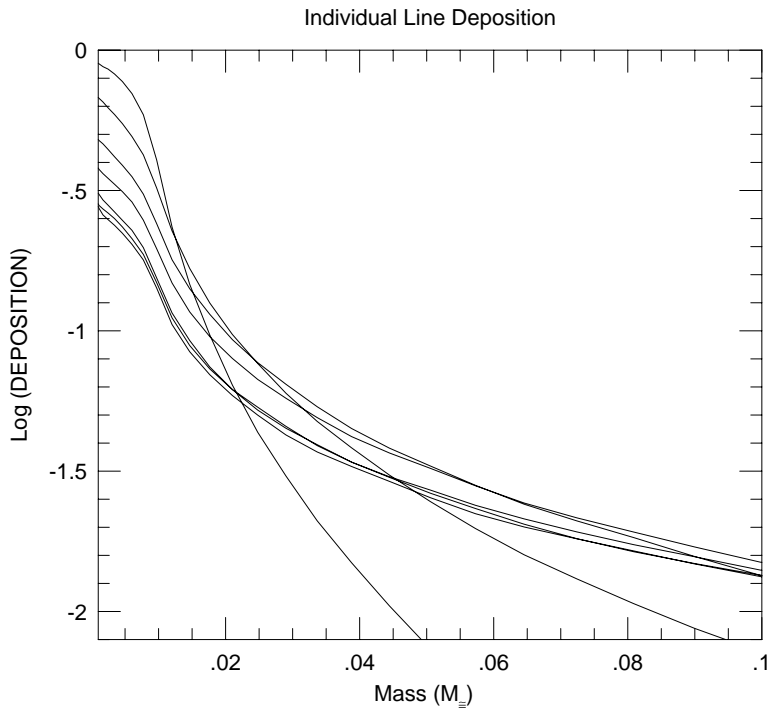

Fig. 2.— The logarithm of the deposition function for mono-energetic sources is shown for a point source ($f_{rad} = 1$ for $M_r < 0.1 M_\odot$) in a constant density, 1 $M_\odot$, model at $t = 30$ days ($\tau_e = 15$). ¿From top to bottom at $M_r = 0$, the deposition corresponds to initial $\gamma$-ray energies $E_\gamma = 0.15$, 0.25, 0.5, 1.0, 3.0, 5.0, and 10.0 MeV. The effective opacity is highest for the lowest initial photon energies, and the deposition is most strongly concentrated near the point of origin as a consequence. The effective opacity is nearly constant for $E_\gamma \gtrsim 2$ MeV. This is reflected by nearly identical deposition functions for this energy range. The W7 model abundances ($Y_e = 0.5$) were used.

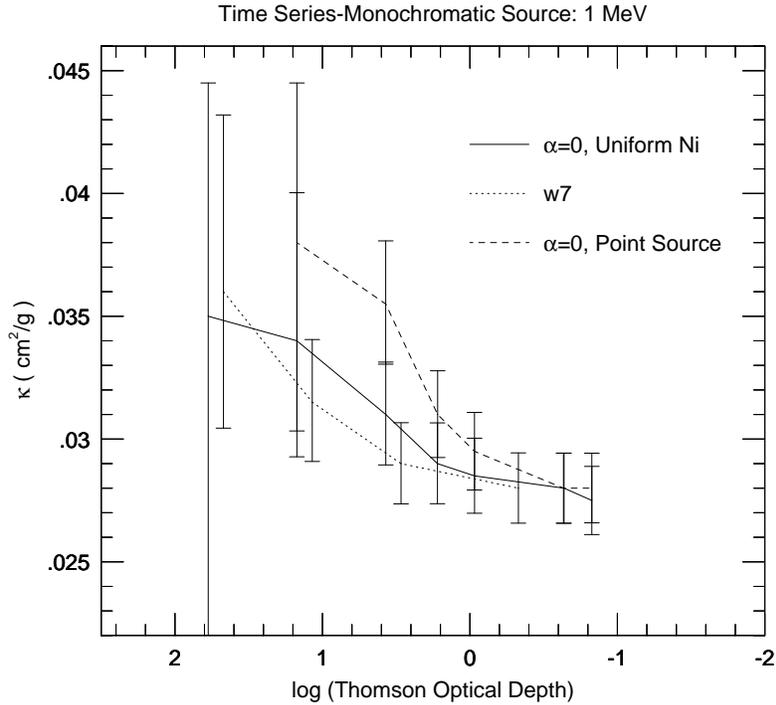

Fig. 3.— The increase in the effective opacity with increasing total optical depth is shown for several models. The models are homologously expanded giving rise to a total Thomson optical depth to the center of the spherically symmetric atmosphere which scales as $\tau_e \propto t^{-2}$. The effective opacity is shown for a 1 MeV source. As $\tau_e \to 0$, $\kappa \to 0.028$ cm$^2$ g$^{-1}$ ($Y_e = 0.5$). Error bars represent the 1-$\sigma$ uncertainty in the best fit effective opacity. As $\tau_e \to \infty$, the deposition function (and hence $Q$ defined by equation 12) is nearly identical for a range of opacities leading to formally large uncertainties at high optical depth.

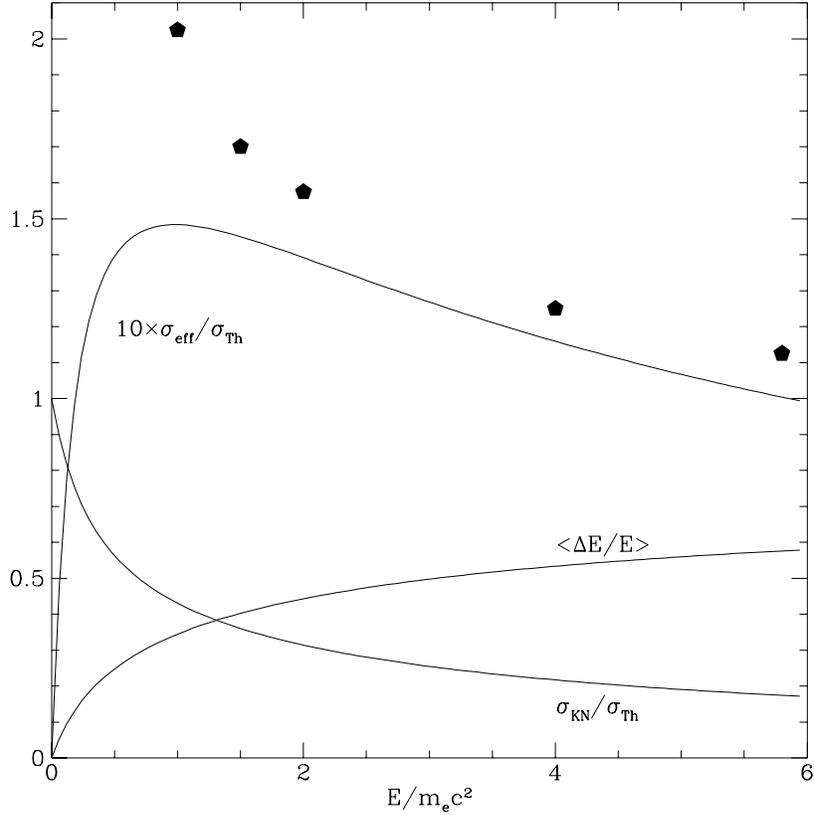

Fig. 4.— The effective cross-section, $\sigma_{eff}$, for single-scattering energy deposition, equation 14, is shown along with the angle-averaged Klein-Nishina cross section, $\sigma_{KN}$, and the angle-averaged fractional energy loss, $\Delta E/E_\gamma$ for a range of incident $\gamma$-ray energies. ($\sigma_{Th}$ is the Thomson cross section. The opacity is simply related to the Compton scattering cross sections as $\kappa = \sigma Y_e/m$.) Symbols denote the effective opacity for the $\alpha = 0$ uniformly distributed source model from Figure 1. For this model, the effective opacity is slightly higher than the analytic solution at all energies due to multiple scattering ($\tau_e \sim 4$ for this model). The larger effective opacity at lower energies is the result of the added opacity due to photoionization.

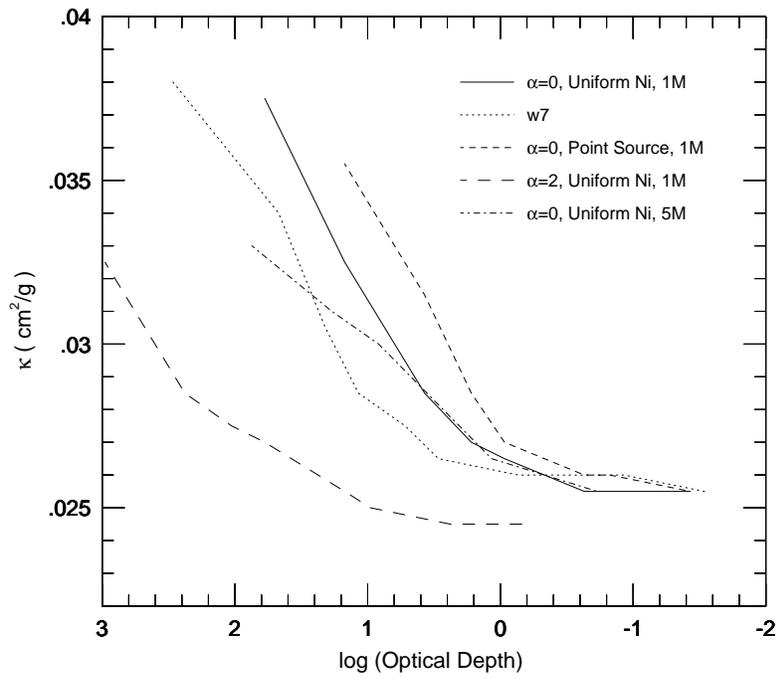

Fig. 5.— The evolving effective opacity for several models is shown as in Figure 3. The full $^{56}$Ni and $^{56}$Co decay spectrum is used here instead of a mono-energetic source as in Figure 3.

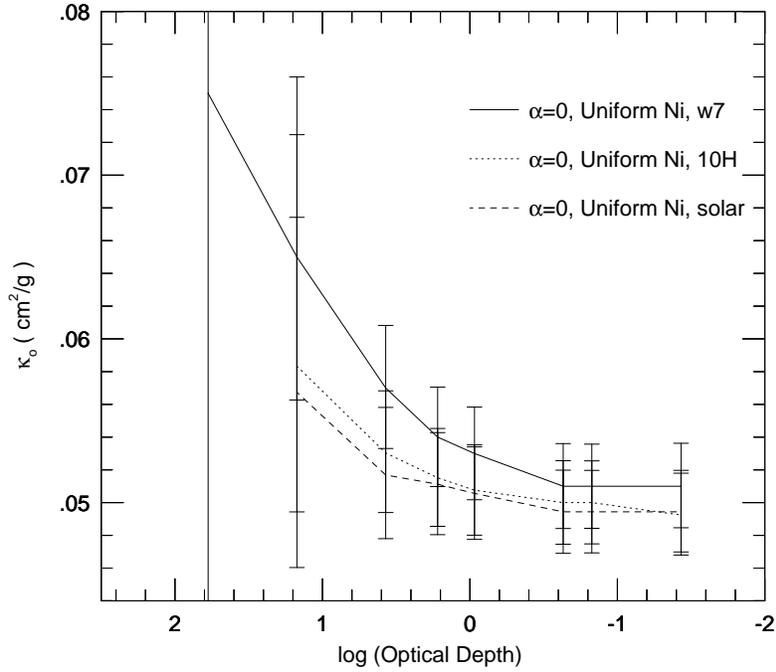

Fig. 6.— The evolving effective opacity for models with different abundances (Table 1) is shown as in Figure 3. In this figure, $\kappa_0 = \kappa_\gamma/Y_e$ is plotted while in previous figures $\kappa_\gamma$ was plotted with $Y_e \equiv 0.5$. The effective opacity, $\kappa_\gamma$, depends on the abundances, to a very good approximation, only through the Compton scattering dependence on the number of available electrons per unit mass.

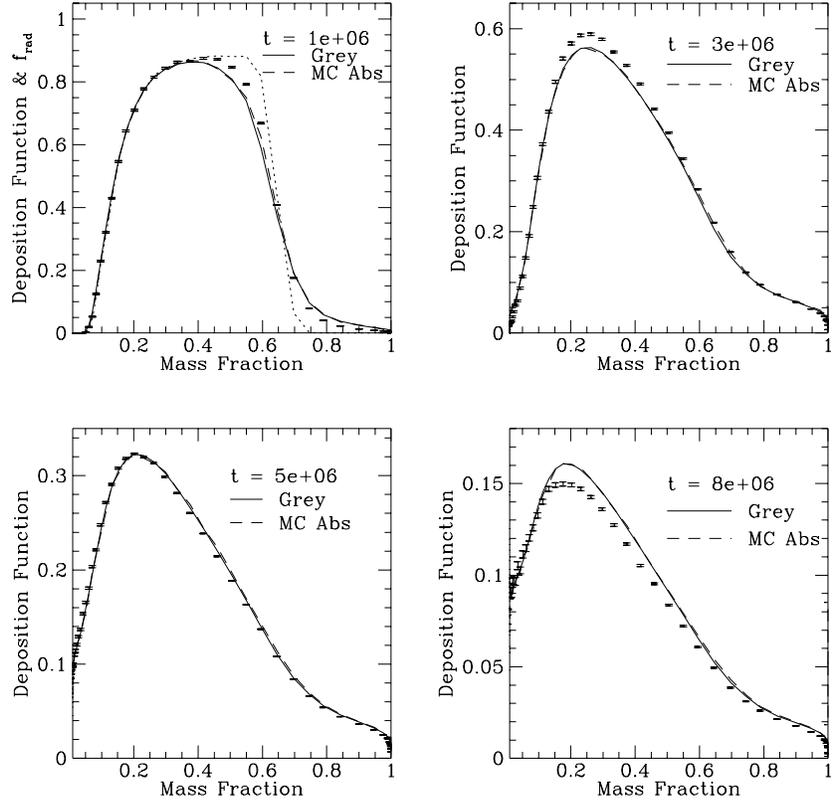

Fig. 7.— Comparison of the Monte Carlo and gray absorption calculations of the deposition function for model W7 of Nomoto, Thielemann, & Yokoi (1984). The dotted line in the top left panel shows the initial mass fraction of radioactive material, $f_{rad}$. The gray absorption calculations have all been performed with $\kappa_\gamma = 0.059 Y_e$ cm$^2$ g$^{-1}$ with $Y_e = 0.5$, which is the optimal value of $\kappa_\gamma$ for $t = 5 \times 10^6$ s after the explosion. Error bars denote the 1-$\sigma$ uncertainties in the Monte Carlo calculations using 200,000 decay events (800,000 for $t = 5 \times 10^6$ s). The *long-dashed* lines denote the Monte Carlo results computed assuming purely absorptive interactions.

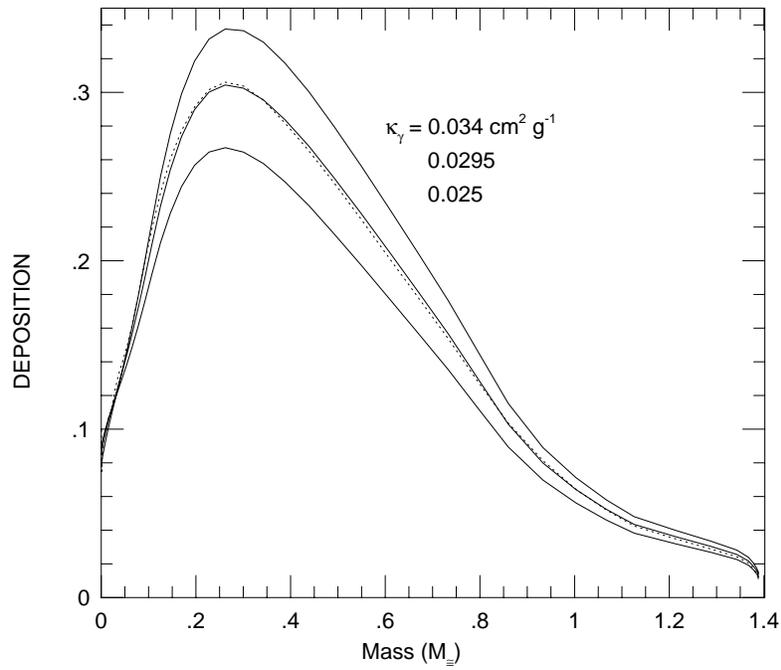

Fig. 8.— The deposition function for model W7 at $t = 5 \times 10^6$ s (see Figure 7) is shown for three values of the effective opacity, $\kappa_\gamma$ (*solid lines*, top to bottom: $\kappa_\gamma = 0.034, 0.0295, 0.025$ cm$^2$ g$^{-1}$). The Monte Carlo deposition function is also shown (*dotted line*). The largest percent error for $\kappa_\gamma = 0.025$ or $0.034$ cm$^2$ g$^{-1}$ is 12%.

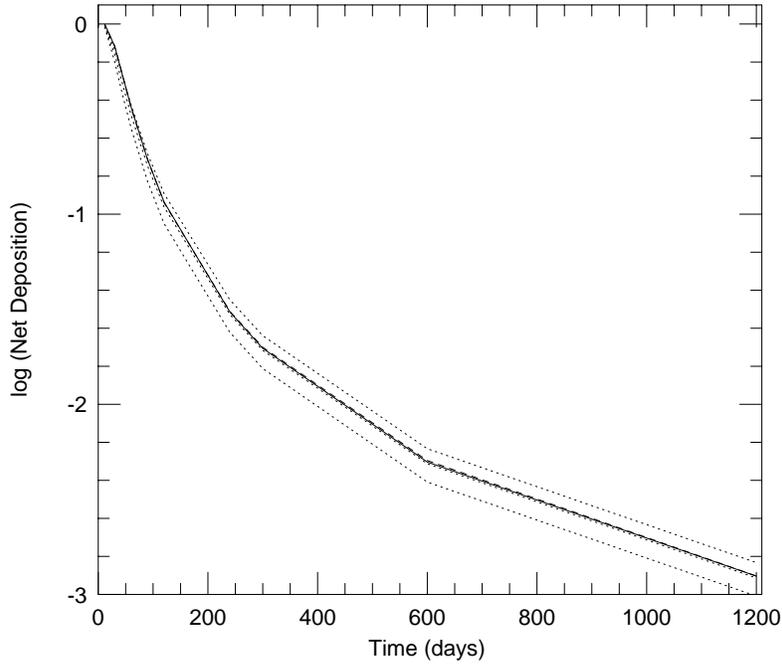

Fig. 9.— The light curve for model W7, defined as $L(t) = \sum_k \Delta M_k d_k(t) \epsilon_{rad} = \epsilon_{rad} D_\gamma M_{rad} \propto D_\gamma$, is shown for the Monte Carlo simulation (*solid line*) and for the gray results using the best-fit value of $\kappa_\gamma$ (*dashed line*) and for three time-independent values of $\kappa_\gamma$ (0.02, 0.025, and 0.03 cm$^2$ g$^{-1}$ *dotted lines*). The luminosities computed using the best-fit value of $\kappa_\gamma$ differs from the Monte Carlo results by less than 2%. Those for $\kappa_\gamma$ = 0.02, 0.025, and 0.03 are in error by less than 22%, 10%, and 18%, respectively. The largest error occurs at $t \sim 90$ days for $\kappa_\gamma$ = 0.02 and 0.025 and at $t = 1200$ days for $\kappa_\gamma = 0.03$. Only the $^{56}$Ni and $^{56}$Co decay spectrum is included. For $t \gtrsim 1000$ days, several less abundant and long-lived radioactive isotopes contribute to the light curve. The contribution from decay positrons is not included.